\documentclass[12pt,twoside]{article}

\usepackage{amsfonts}
\usepackage{amsmath}
\usepackage{cite}
\usepackage{epsfig}
\usepackage{a4}

\newcommand{\I}{\textrm{i}}        % imagin"are Einheit
\newcommand{\E}{\mathrm{e}}        % e Eulersche Zahl
\newcommand{\D}{\mathrm{d}}       % Differential

\newcommand{\ch}{\operatorname{ch}}
\newcommand{\sh}{\operatorname{sh}}

\newcommand{\XXZ}{{X\!X\!Z}}
\newcommand{\scaling}{continuum\ }
\newcommand{\Scaling}{Continuum\ }

\begin{document}

\thispagestyle{empty}
\parindent0cm

\begin{center}

{\Large {\bf A note on the spin-$\mathbf{\frac{1}{2}}$ XXZ chain concerning its relation to the Bose gas\\}}

\vspace{7mm}

{\large Alexander Seel\footnote[1]{e-mail:
 alexander.seel@itp.uni-hannover.de}, 
Tanaya Bhattacharyya\footnote[2]{e-mail: bhattach@lmpt.univ-tours.fr},\\ 

Frank G\"ohmann\footnote[3]{e-mail: goehmann@physik.uni-wuppertal.de}
and Andreas Kl\"{u}mper\footnote[4]{e-mail: kluemper@physik.uni-wuppertal.de}\\

\vspace{5mm}

$^1$Institut f\"ur Theoretische Physik, Leibniz Universit\"at Hannover,\\
30167 Hannover, Germany\\[2ex]

$^2$Laboratoire de Math\'ematiques et Physique Th\'eorique\\
CNRS/UMR 6083, Universit\'e de Tours,
Parc de Grandmont,\\ 37200 Tours, France\\[2ex]

$^{3,4}$Fachbereich C -- Physik, Bergische Universit\"at Wuppertal,\\
42097 Wuppertal, Germany\\}

\vspace{20mm}

{\large {\bf Abstract}}

\end{center}

\begin{list}{}{\addtolength{\rightmargin}{10mm}
               \addtolength{\topsep}{-5mm}}
\item

  By considering the one-particle and two-particle scattering data of the
  spin-$\frac{1}{2}$ Heisenberg chain at $T=0$ we derive a \scaling limit
  relating the spin chain to the 1D Bose gas. Applying this limit to the quantum
  transfer matrix approach of the Heisenberg chain we obtain expressions for
  the correlation functions of the Bose gas at arbitrary temperatures. \\[2ex]
% 05.30.-d quantum statistical mechanics
%75.10.Pq Spin chain models
%02.30.Ik integrable Systems
%02.30.-f Function theory, analysis
\end{list}

\clearpage

\section{Introduction}

Bethe's famous investigation on the spectrum of the Heisenberg chain in 1931
\cite{Bethe31} was performed by a method later termed Bethe's ansatz and was
found, after suitable modifications, to be applicable to other many-particle
systems, e.g. to the repulsive Bose gas \cite{LiLi63} and the Hubbard model\cite{LiWu68}.\\

The algebraic reason for the exact solvability of these models was found in
the Yang-Baxter relation \cite{Baxter82} and allowed for rather powerful solution
techniques on the basis of the quantum inverse scattering method
\cite{SkTaFa79}. The eigenstates of the Heisenberg Hamiltonian are obtained
from a trivial vacuum by the action of certain operators satisfying the
Yang-Baxter algebra, an algebra with quadratic relations and structure
coefficients given by the $R$-matrix of the six-vertex model. Within this
algebraic Bethe ansatz effective algebraic formulae for scalar products
\cite{Slavnov89} and norms \cite{Korepin82} were derived.\\

As a further development, Kitanine et al. \cite{KMT99a} were able to express
local operators of the $\XXZ$ chain in terms of the algebra and to calculate
static correlation functions at $T=0$ by evaluating scalar products. Thus they
rederived the multiple integral representations for correlation functions by
Jimbo et al. \cite{JMMN92,JiMi96} and generalized them to finite longitudinal
magnetic fields \cite{KMT00,KMST02a}.\\[-1.5ex]

The next logical step was to introduce finite temperatures with a first
application to the generating function \cite{IzKo84} of the $zz$-correlation. This was achieved by
G\"ohmann et al. in \cite{GoKlSe04}, where the algebraic structure of
\cite{Slavnov89,KMST02a} underlying the treatment of ground state correlation
functions was combined with the functional approach
\cite{Kluemper92,Kluemper93,Kluemper04} for the quantum transfer matrix (QTM).
This object describes an auxiliary spin chain whose leading eigenvalue is
nothing but the partition function of the $\XXZ$ chain and the eigenvector
encodes all thermal correlations.\\

In the latest development, with respect to the Heisenberg chain, Sakai
\cite{Sakai07} succeeded in deriving a multiple integral representation for
the generating function of dynamical correlations for finite temperatures.\\

An alternative method to the QTM approach, and actually a more traditional
treatment of finite temperatures is the so-called thermodynamical Bethe ansatz
(TBA) introduced by Yang and Yang \cite{YaYa69} for the 1D Bose gas. Knowing
the spectrum in terms of the Bethe ansatz \cite{LiLi63}, a combinatorial
expression for the entropy and a minimization of a free energy functional lead
to the thermodynamics. For the Bose gas this approach yields an integral
expression for the thermodynamical potential in terms of an auxiliary function
to be determined from a non-linear integral equation. The same approach, however, for
the Heisenberg model yields many more integral equations for the
bound states of magnons underlying the 'string hypothesis'.\\

The goal of this paper is the computation of correlation functions for the
integrable Bose gas at finite temperature by use of technical tools developed
for the spin-1/2 Heisenberg chain. Unfortunately, the direct application of
the QTM approach fails as it is a concept that is most natural for lattice
systems. Here we apply these techniques to a lattice system containing the
Bose gas as a suitable continuum limit. This is where the anisotropic spin-1/2 
Heisenberg chain and the Bose gas meet.\\

In section 2 we review the Bethe ansatz solution of the spin-1/2 Heisenberg
chain with $\XXZ$ anisotropy. In section 3 we identify the continuum limit
taking the one-particle spectral and the two-particle scattering properties of
the Heisenberg chain to those of the Bose gas. In section 4 we demonstrate how
to rederive the famous TBA equations of the Bose gas from thermodynamical
equations of the Heisenberg chain in the continuum limit. In section 5, a
useful generating function for correlations is identified and results for this
generating function in case of the Heisenberg chain are quoted from the
literature. These expressions are then translated in section 6 into (novel)
results for the 1D Bose gas by use of the continuum limit. This completes our
demonstration on how to obtain correlation functions for continuum models from
those of lattice systems.

%%%%%%%%%%%%%%%%%%%%%%%%%%%%%%%%%%%%%%%%%%%%%%%%%%%%%%%%%%%%%%%%%%%%

\section{The Heisenberg Chain}
The Hamiltonian of the anisotropic ${\XXZ}$ chain on $L$ lattice sites exposed
to an external longitudinal magnetic field $h$,
\begin{equation} \label{XXZh}
   H_{L} = J \sum_{j=1}^L
                \bigg[ \sigma_{j-1}^x\sigma_{j}^x
		       + \sigma_{j-1}^y\sigma_{j}^y
		       + \Delta (\sigma_{j-1}^z\sigma_{j}^z - 1) \bigg]
		       - \frac{h}{2} \sum_{j=1}^L \sigma_j^z \quad ,
\end{equation}
is closely related to the transfer matrix of a six-vertex model with local
Boltzmann weights encoded in the $R$-matrix
\begin{equation} \label{Rmatrix}
R(\lambda,\mu) =
  \begin{pmatrix}
     1 & 0 & 0 & 0 \\
     0 & b(\lambda,\mu) & c(\lambda,\mu) & 0 \\
     0 & c(\lambda,\mu) & b(\lambda,\mu) & 0 \\
     0 & 0 & 0 & 1
  \end{pmatrix} \qquad , \qquad
  \begin{gathered}
     b(\lambda,\mu) = \frac{\sh(\lambda-\mu)}
                           {\sh(\lambda-\mu+\eta)}\\
     c(\lambda,\mu) = \frac{\sh \eta}{\sh(\lambda-\mu+\eta)}
  \end{gathered} \quad  .
\end{equation}

The coupling strength of the Hamiltonian is denoted by $J$,
$\Delta=\ch\eta$ is the anisotropy and $\sigma_j^\alpha$, $\alpha=x,y,z$
denote the usual Pauli matrices.  The eigenvalue problem of the
transfer matrix of the six-vertex model can be solved for instance in the framework of
the algebraic Bethe ansatz, see e.g. \cite{KBIBo}. The energies
$E_M$ of \eqref{XXZh} and momenta $\Pi_M$ are given by
\begin{equation}
{E_M = 2J\sh\eta \,\frac{\partial\ln\Lambda_M}{\partial\lambda}({\eta}/{2})
      - \big({L}/{2} - M\big)h \;\;\; , \;\;\; \Pi_M = -\I \ln\Lambda_M
      ({\eta}/{2}) }
\end{equation}
where the transfer matrix eigenvalue $\Lambda_M(\lambda)$ is obtained from
\begin{equation}
{\Lambda_M(\lambda) = 
      \Big[\prod_{l=1}^M \frac{\sh(\lambda-\lambda_l-\eta)}{\sh(\lambda-\lambda_l)}\Big] +
      \Big(\frac{\sh(\lambda-\eta/2)}{\sh(\lambda+\eta/2)}\Big)^L
      \Big[\prod_{l=1}^M \frac{\sh(\lambda-\lambda_l+\eta)}{\sh(\lambda-\lambda_l)}\Big] }
\end{equation}
provided that the Bethe equations
\begin{equation} \label{Betheeqn}
  \Big(\frac{\sh(\lambda_j-\eta/2)}{\sh(\lambda_j+\eta/2)}\Big)^L =
  \Big[\prod_{\substack{l = 1 \\l \not= j}}^M
  \frac{\sh(\lambda_j-\lambda_l-\eta)}{\sh(\lambda_j-\lambda_l+\eta)}\Big]
  \quad , \quad j=1,\ldots,M
\end{equation}
for the $M$ Bethe roots $\lambda_j$ are satisfied.\\

The above listed Bethe ansatz equations allow for the computation of the
entire spectrum of the system. For identifying a suitable scaling limit, in which the
$\XXZ$ lattice system turns into the Bose gas in the continuum, it is sufficient to
consider the one-particle properties and the two-particle scattering data.  In
particular, we are going to identify the ferromagnetic vacuum with the Fock vacuum for Bosons, and flipped spins on the ferromagnetic background are
considered as Bosons.

In the one-magnon sector $(M=1)$ the energy-momentum dispersion simply reads (with $E_0 = -hL/2$)
\begin{equation}\label{dispersion}
E_1-E_0 = \frac{2J\sh^2\eta}{\sh(\lambda_1+\eta/2)\sh(\lambda_1-\eta/2)} + h \quad , \quad
\Pi_1 = \I \ln\frac{\sh(\lambda_1 -\eta/2)}{\sh(\lambda_1+\eta/2)}
\end{equation}
and can be considered as some one-particle excitation energy $E$ over the
vacuum \mbox{$(M=0)$} with a chemical potential $\mu$,
\begin{math}
E_1 - E_0 =: E-\mu
\end{math}.

%%%%%%%%%%%%%%%%%%%%%%%%%%%%%%%%%%%%%%%%%%%%%%%%%%%%%%%%%%%%%%%%

\section{\Scaling Limit}

In the desired \scaling limit, the ferromagnetic point $\Delta=-1$ is
approached from the region $|\Delta|<1$ in a well defined way. Therefore we
specify the anisotropy of the interaction as $\Delta=\ch\eta$ with $\eta =\I\pi
-\I\varepsilon$. Inserting this into the Bethe equations \eqref{Betheeqn} 
as well as the reparametrization $\lambda_j = x_j/L$ of the Bethe roots leads to
\begin{align}
\label{LHS}
LHS &= \Big(\frac{\sh(\lambda_j-\eta/2)}{\sh(\lambda_j+\eta/2)}\Big)^L 
    \approx\Big(\frac{1+\I\varepsilon x_j/2L}{1 - \I\varepsilon x_j/2L}\Big)^L
    \to \E^{\I\varepsilon x_j} \\[.5\baselineskip]
RHS &= \Bigg[\prod_{\substack{l = 1 \\l \not= j}}^M \frac{\sh(\lambda_j-\lambda_l-\eta)}{\sh(\lambda_j-\lambda_l+\eta)}\Bigg] 
  \approx \Bigg[\prod_{\substack{l = 1 \\l \not= j}}^M 
  \frac{x_j - x_l +\I L \varepsilon }{x_j - x_l -\I L \varepsilon}\Bigg]
\end{align}
where we have used the formula $\lim_{n\to\infty}(1+x/n)^n = \E^x$ for ``LHS''
and anticipated $L\to\infty$ and $\varepsilon x_j$ finite. Introducing a
lattice spacing $\delta$ such that the chain has the physical length
$\ell=\delta L$, the rescaling $\varepsilon x_j = \ell \nu_j$ reveals the
Bethe equations to be exactly those of the Bose gas \cite{LiLi63},
\begin{equation}
  \E^{\I \ell \nu_j} = \Bigg[\prod_{\substack{l=1\\l\not=j}}^M 
     \frac{\nu_j - \nu_l + \I c}{\nu_j - \nu_l - \I c}\Bigg]
 \quad , \quad j=1,\ldots,M \quad ,
\end{equation}
where the repulsion strength can be identified as $c=\varepsilon^2/\delta$. Furthermore, the single-particle energy-momentum dispersion is
\begin{equation} \label{dispersion2}
E = 2J (\delta \nu)^2
\end{equation}
with mass $m_B=1/4J\delta^2$ (implying $J\to\infty$) and a chemical potential
$\mu = 2J\varepsilon^2 -h$ directly connected to the magnetic field. For convenience we set $m_B=1/2$ as Lieb and Liniger \cite{LiLi63} already did.\\

%%%%%%%%%%%%%%%%%%%%%%%%%%%%%%%%%%%%%%%%%%%%%%%%%%%%%%%%%

{\bf Parameters for the Spin-$\mathbf{\frac{1}{2}}$ Heisenberg Chain and the Bose Gas}\\[.5\baselineskip]
Let us recapitulate the parameters of the spin chain and the Bose gas with
lattice spacing $\delta$ and the anisotropy parametrized by
$\Delta=-\cos\varepsilon$, $\varepsilon\ll 1$:

\begin{center}
\begin{tabular}{l|l}
$\XXZ$ chain & Bose gas \\[.2ex]
\hline \\[-2ex]
interaction strength $J > 0$ & particle mass $m_B=1/4J\delta^2$\\
\# lattice sites $L$, lattice constant $\delta$ & physical length $\ell=L\delta$\\
magnetic field $h > 0$ & chemical potential $\mu =  2J\varepsilon^2 - h$\\
anisotropy $\Delta=\varepsilon^2/2 - 1$ & repulsion strength $c=\varepsilon^2/\delta$
\end{tabular} 
\end{center}

This allows for a well-defined \scaling limit for the five lattice parameters
while keeping the four continuum parameters fixed. In the next section we are
going to apply the above \scaling limit, established at $T=0$, to the finite
temperature formalism of the Heisenberg chain.

%%%%%%%%%%%%%%%%%%%%%%%%%%%%%%%%%%%%%%%%%%%%%%%%%%%%%%%%%%%%%%%%%%%%%%%%

\section{Non-linear Integral Equation and Free Energy}
\label{NonlinFreeEn}
\begin{figure}
\begin{center}
\epsfig{file=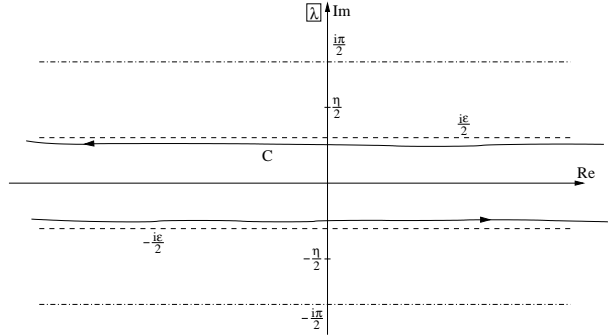,width=8cm}
\caption{\label{CKont} Depiction of the canonical contour $C$ surrounding the real axis in
  counterclockwise manner within the strip $-\varepsilon/2 < \operatorname{Im}
  \lambda < \varepsilon/2$.}
\end{center}
\end{figure}

The free energy $F_{\XXZ}/L$ per lattice site for the Hamiltonian \eqref{XXZh}
follows within the quantum transfer matrix approach \cite{Kluemper04} as
\begin{equation}\label{free2}
\frac{F_\XXZ}{L} = -\frac{h}{2} - 
   T\int_C \frac{\D\omega}{2\pi\I} \frac{\sh\eta}{\sh\omega\sh(\omega+\eta)} 
   \ln(1+\mathfrak{a}(\omega))\quad,
\end{equation}
where the auxiliary function $\mathfrak{a}(\lambda)$ is calculated from the
non-linear integral equation
\begin{multline}\label{inteqa}
     \ln \mathfrak{a}(\lambda) = -\frac{h}{T}
        +  \frac{2J \sh^2(\I\varepsilon)}{T\sh \lambda
	        \sh(\lambda -\I\varepsilon)}   \\[.5\baselineskip]
        + \int_C \frac{\D\omega}{2\pi\I}
	   \frac{\sh(2\I\varepsilon)}{\sh(\lambda - \omega +\I\varepsilon)
	   \sh(\lambda - \omega -\I\varepsilon)} \ln (1 + \mathfrak{a}(\omega))  \quad .\quad
\end{multline}
Here the temperature is denoted by $T$ and the external magnetic field $h$ in the
\scaling limit (with the fully polarized state in positive direction
corresponding to the vacuum) takes positive values.
In the integrals above, the contour $C$ is a rectangular path centered around
zero.  In the critical regime under consideration, with $\eta=\I\pi
-\I\varepsilon$, its height is restricted by $\varepsilon$ and the width tends
to infinity as depicted in figure \ref{CKont}.\\

Applying the low-$T$ limit (note that $J/T\to\infty$) to the integral equation
the contribution from the upper line of the contour at
$\omega=\I\varepsilon/2$ vanishes and one can restrict both the free argument
$\lambda$ and the integration variable $\omega$ to the lower part of the
remaining contour. A shift of this line to imaginary part $-\eta/2$ without
crossing any poles of the inhomogeneity results in the line integral
\begin{multline} \label{hilfsfunkReel}
\ln \mathfrak{a}(\lambda-\eta/2) = -\frac{h}{T}
   - \frac{2J\sh^2\eta}{T\sh(\lambda-\eta/2)\sh(\lambda+\eta/2)}  \\[.5\baselineskip]
 +\int_\mathbb{R} \frac{\D\omega}{2\pi\I} 
\frac{\sh(2\I\varepsilon)\ln (1 + \mathfrak{a}(\omega-\eta/2))}
          {\sh(\lambda - \omega +\I\varepsilon)\sh(\lambda - \omega -\I\varepsilon)} \quad
\end{multline}
over the real axis. Observing that in the \scaling limit the inhomogeneity of
\eqref{hilfsfunkReel} is nothing but the one-particle excitation energy with
some chemical potential, cf. \eqref{dispersion} and \eqref{dispersion2}, the
new auxiliary function reads
\begin{equation}\label{auxneu}
  \ln \widetilde{\mathfrak{a}}(\nu) = -\frac{E-\mu}{T} 
  + \int_{\mathbb{R}}\frac{\D w}{2\pi} \frac{2c}{(\nu-w)^2+c^2} 
\ln (1 + \widetilde{\mathfrak{a}}(w))\quad .
\end{equation}

For the free energy we proceed in the same way. First, we note that the Bose
gas grand canonical potential per unit length $\phi$ corresponds to the free
energy of the Heisenberg chain per lattice constant $\delta$, in detail
$\phi=(F_{\XXZ}/L + h/2)/\delta$. Second, starting from the integral
representation \eqref{free2} at low temperatures only the lower part of the
contour contributes. Shifting this integration line to imaginary part
$-\eta/2$ without crossing poles from the kernel we get in the \scaling limit
\begin{equation}\label{phi}
 \phi=  - T \int_\mathbb{R}\frac{\D w}{2\pi} \ln (1 + \widetilde{\mathfrak{a}}(w)) \quad .
\end{equation}
Looking up the TBA results \cite{YaYa69}
of Yang and Yang provides the same integral representations \eqref{auxneu} and
\eqref{phi} describing the thermodynamics.

%%%%%%%%%%%%%%%%%%%%%%%%%%%%%%%%%%%%%%%%%%%%%%%%%%%%%%%%%%%%%%%%%%%%%%%%%%%%%%%%

\section{Finite Temperature Correlation Functions}
The repulsive 1D Bose gas is captured by the quantum non-linear Schr\"odinger
equation. The Hamiltonian for a system of the physical length $\ell$ then
reads
\begin{equation}
H_B = \int_0^\ell\!\!\D z \Big[\partial_z\psi^\dagger(z)\partial_z\psi(z)
+ c\, \psi^\dagger(z)\psi^\dagger(z)\psi(z)\psi(z)
- \mu\, \psi^\dagger(z)\psi(z)\Big] \quad ,
\end{equation}
where the operators $\psi^\dagger$ and $\psi$ are canonical Bose fields and the mass of the Bosons is set to $m_B=1/2$ for convenience.
Aiming at the number of particles in the interval $[0,x]$ the corresponding
operator $Q_1(x)$ has the explicit representation
\begin{equation} \label{Q1Op}
Q_1(x) = \int_0^x\!\!\D z\,\psi^\dagger(z)\psi(z)
\end{equation}
and the ground state expectation value of $\exp({\varphi Q_1(x)})$ can be used to compute the density-density correlation function \cite{IzKo84} reading
\begin{equation}
\left.\langle \psi^\dagger(x)\psi(x)\psi^\dagger(0)\psi(0) \rangle 
= \frac{1}{2} \frac{\partial^2}{\partial \varphi^2}  
  \frac{\partial^2}{\partial x^2}\langle \E^{\varphi Q_1(x)}\rangle\right|_{\varphi=0} \quad .
\end{equation}

Recently Kitanine et al. \cite{KKMST07} derived a multiple-integral
representation for the density-density correlation function for the Bose gas
at $T=0$ -- similar to the generating function $\exp(\varphi Q_{1,m}) $ of the
$zz$-correlation for the $\XXZ$ chain. By identifying the operator $Q_1(x)$ with $Q_{1,m}$ involving $m$ consecutive sites the definition for the spin chain reads
\begin{equation} \label{Q}
 Q_{1,m} = \frac{1}{2} \sum_{j=1}^m (1 - \sigma_j^z) \quad .
\end{equation}

We already showed \cite{GoKlSe04} that within the quantum transfer matrix
approach the thermal expectation value $\langle \textrm{e}^{\varphi Q_{1,m}}
\rangle _T$ has the multiple integral representation
\begin{align}  \notag
{\langle \textrm{e}^{\varphi Q_{1,m}}\rangle _T} = \sum_{n=0}^m \frac{1}{(n!)^2}
 &\Bigg[\prod_{j=1}^n\int_C \frac{\D\omega_j}{2\pi\I}
 \frac{{\mathfrak{a}}(\omega_j)}{1+{\mathfrak{a}}(\omega_j)}
 \Big(\frac{\sh \omega_j}{\sh(\omega_j+\eta)}\Big)^m\Bigg] \\[.5\baselineskip]  \notag
 &\Bigg[\prod_{j=1}^n \int_\Gamma \frac{\D z_j}{2\pi\I}
 \Big(\frac{\sh(z_j+\eta)}{\sh z_j}\Big)^m\Bigg]
 \Bigg[\prod_{j,k=1}^n
 \frac{\sh(\omega_j-z_k+\eta)}{\sh(z_j-z_k+\eta)}\Bigg]\\[.5\baselineskip]
\label{erzeugfunk}
 &\,\,\,\operatorname{det}_n M(\omega_j,z_k) 
 \operatorname{det}_n G(\omega_j,z_k) &
\end{align}

with the $n \times n$ matrices
\begin{multline} \label{MMat} {M}(\omega_j,z_k) = -\frac{\sh
    \eta}{\sh(\omega_j-z_k)\sh(\omega_j-z_k-\eta)}\,
  \Bigg[\prod_{l=1}^n
  \frac{\sh(\omega_j-z_l-\eta)}{\sh(\omega_j-\omega_l-\eta)}\Bigg]\\[.5\baselineskip]
  - \frac{\sh \eta\,\,\textrm{e}^\varphi}
  {\sh(\omega_j-z_k)\sh(\omega_j-z_k+\eta)} \, \Bigg[\prod_{l=1}^n
  \frac{\sh(\omega_j-z_l+\eta)}{\sh(\omega_j-\omega_l+\eta)}\Bigg] \quad
\end{multline}

and $G(\omega_j,z_k)$, where $G(\lambda,z)$ is defined by the linear integral
equation
\begin{multline} \label{Gfunk}
G(\lambda,z) = -\frac{\sh \eta}{\sh(\lambda-z)\sh(\lambda-z+\eta)}\\[.5\baselineskip]
- \int_C \frac{\D\omega}{2\pi\I}
  \frac{\sh(2\eta)}
  {\sh(\lambda-\omega+ \eta) \sh(\lambda-\omega-\eta)}
  \frac{\mathfrak{a}(\omega) G(\omega,z)}{1 + \mathfrak{a}(\omega)}\quad .
\end{multline}

The integration path $\Gamma$ surrounds the origin in counterclockwise manner
and has to be enclosed by the canonical contour $C$ occurring already in
\eqref{inteqa}.

%%%%%%%%%%%%%%%%%%%%%%%%%%%%%%%%%%%%%%%%%%%%%%%%%%%%%%%%%%%%%%%%%%%%%%%%%%%%%

\section{Density-Density Correlations of the Bose Gas}
For applying the Bose limit to the integral representation \eqref{erzeugfunk}, let
us note that the contour $C$ can be chosen to consist of two
horizontal lines with imaginary parts $\pm\I\varepsilon/2$, and $\Gamma$
similarly, but inside $C$.\\

Proceeding with the low-$T$ limit of \eqref{erzeugfunk}, only the lower
integration lines of $C$ remain due to the factors
$\mathfrak{a}/(1+\mathfrak{a})$. As the upper line of the path $\Gamma$ is
then no longer bounded by the contour $C$ it can be shifted to imaginary part
$\I\varepsilon/2$. Now in \eqref{erzeugfunk}, the $z_j-$integral over this line
vanishes in the continuum limit as the factor 
\begin{equation}
\left(\frac{\sh(z_j+\eta)}{\sh z_j}\right)^m\to
\left(-\frac{v_j-\I c/2}{v_j+\I c/2}\right)^{x/\delta}
\end{equation}
in the integrand becomes rapidly oscillating as $m={x/\delta}\to\infty$ for fixed $x$.\\

Like in {section \ref{NonlinFreeEn}}, a simultaneous shift of all remaining (lower)
integration lines to imaginary part $-\eta/2$ results in
\begin{align} \notag \langle \textrm{e}^{\varphi Q_{1,m}} \rangle _T =
  \sum_{n=0}^m \frac{1}{(n!)^2} \Bigg[\prod_{j=1}^n
\int\displaylimits_\mathbb{R}
  \frac{\D\omega_j}{2\pi\I}
  \frac{{\mathfrak{a}}(\omega_j-\eta/2)}{1+{\mathfrak{a}}(\omega_j-\eta/2)}
  \Big(\frac{\sh (\omega_j-\eta/2)}{\sh(\omega_j+\eta/2)}\Big)^m\Bigg]&  \\[.5\baselineskip]
  \notag \Bigg[\prod_{j=1}^n 
\int\displaylimits_{\phantom{x}\mathbb{R}+\I0} \frac{\D
    z_j}{2\pi\I} \Big(\frac{\sh(z_j+\eta/2)}{\sh( z_j-\eta/2)}\Big)^m\Bigg]
  \Bigg[\prod_{j,k=1}^n \frac{\sh(\omega_j-z_k+\eta)}{\sh(z_j-z_k+\eta)}
\Bigg]\\[.5\baselineskip]
 \label{preform}
 \operatorname{det}_n M(\omega_j-\eta/2,z_k-\eta/2) 
 \operatorname{det}_n G(\omega_j-\eta/2,z_k-\eta/2)&
\end{align}
for the Bose limit. Suitably reparametrising the variables of integration
we find the final form
\begin{align} \notag
{\langle \textrm{e}^{\varphi Q_{1}}\rangle _T} = \sum_{n=0}^\infty \frac{1}{(n!)^2}
&\Bigg[ \prod_{j=1}^n \int\displaylimits_\mathbb{R}
 \frac{\D w_j \widetilde{\mathfrak{a}}(w_j)}{1+\widetilde{\mathfrak{a}}(w_j)}
 \int\displaylimits_{\mathbb{R}+\I 0}\!\!\!\!\frac{\D p_j}{2\pi}\E^{-\I (p_j-w_j)x}
 \Bigg]\\[.5\baselineskip]\notag
& \Bigg[\prod_{j,k=1}^n \frac{w_j - p_k -\I c}{p_j - p_k -\I c}\,
    \frac{p_k - w_j -\I c}{w_j - w_k -\I c}\Bigg]\\[.5\baselineskip] 
&\,\,\,\,\operatorname{det}_n \widetilde{M}(w_j,p_k) 
 \operatorname{det}_n \rho(w_j,p_k)  \quad .
\end{align}
As above, we set the physical length of the $m$ consecutive sites to be
$x=m\delta$ and the limit $\delta\to 0$ led to $m\to\infty$ and to the infinite
series on the RHS of the last equation. 
The matrix
\begin{multline}
\widetilde{M}(w_j,p_k) = -\frac{c}{(w_j-p_k)(w_j-p_k+\I c)} \\[.5\baselineskip] 
- \frac{c\,\,\E^\varphi}{(w_j-p_k)(w_j-p_k-\I c)} 
\Bigg[\prod_{l=1}^n\frac{p_l - w_j +\I c}{w_j - p_l +\I c}\,
\frac{w_j - w_l +\I c}{w_l - w_j +\I c} \Bigg] \quad 
\end{multline}
follows from \eqref{MMat} and as
a direct consequence of \eqref{Gfunk} the density $\rho(w,p)$ is 
the solution of the linear integral equation
\begin{equation}
2\pi \rho(\nu,p) = - \frac{c}{(\nu-p)(\nu-p-\I c)} 
+ \int\displaylimits_{\mathbb{R}}\frac{2c\,\,\D w}{(\nu-w)^2 + c^2}
  \frac{\widetilde{\mathfrak{a}}(w)\rho(w,p)}{1+\widetilde{\mathfrak{a}}(w)} \quad .
\end{equation}
By use of means rather different from
those of \cite{KKMST07} we found a new way to derive correlations of the Bose
gas with even wider applicability, namely to the finite temperature case.

 %%%%%%%%%%%%%%%%%%%%%%%%%%%%%%%%%%%%%%%%%%%%%%%%%%%%%%%%%%%%%%%%%%%%%%

\section{Conclusion}

On the level of the Bethe ansatz equations for $T=0$ we showed how to relate
the $\XXZ$ chain near the ferromagnetic point to the 1D repulsive Bose gas.
Applying this to the QTM approach reproduces the known TBA thermodynamics of
Bose particles. Then, and most importantly, we were able to derive a multiple
integral representation for the generating function of the
density-density correlations of the Bose gas, valid for arbitrary temperatures.\\

Unfortunately, the multiple integral representation is an infinite series with an infinite number of integrals to be calculated. In this respect it resembles the Fredholm determinant representation \cite{KoKo97,KKS97} but avoids dual quantum fields. Apart from the difficulties lying in the infinite series the derivation of explicit results on the temperature dependence of the correlation lengths is feasible by considering the next-to-leading eigenvalues of the quantum transfer matrix.\\

Furthermore it is interesting how to interpret and how to treat other correlation functions of the $\XXZ$ chain in the considered continuum limit.\\

%%%%%%%%%%%%%%%%%%%%%%%%%%%%%%%%%%%%%%%%%%%%%%%%%%%%%%%%%%%%%%%%%%%%%

{\bf Acknowledgement.} The authors would like to thank C. Bircan for
interesting discussions. AS gratefully acknowledges financial support by the
German Science Foundation under grant number Se~1742~1-1. TB gratefully acknowledges support from Alexander von Humboldt Foundation as this work was mainly done during her stay in the University of Wuppertal as a Humboldt Fellow.

\bibliographystyle{amsplain}
% \bibliography{qtm}     

%%%%%%%%%%%%%%%%%%%%%%%%%%%%%%%%%%%%%%%%%%%%%%%%%%%%%%%%%%%%%%%%%%%%%%%%
% ab hier Datei ***.bbl von BibTex :
\providecommand{\bysame}{\leavevmode\hbox to3em{\hrulefill}\thinspace}
\providecommand{\MR}{\relax\ifhmode\unskip\space\fi MR }
\providecommand{\MRhref}[2]{%
  \href{http://www.ams.org/mathscinet-getitem?mr=#1}{#2}
}
\providecommand{\href}[2]{#2}

%%%%%%%%%%%%%%%%%%%%%%%%%%%%%%%%%%%%%%%%%%%%%%%%%%%%%%%%%%%%%%%%%%%%%%%%%

\end{document}